\documentstyle[aps,multicol,prl,psfig,epsf]{revtex}

\begin{document}
\title
{
Dynamical pinning and non-Hermitian mode transmutation in the
Burgers equation
}
\author{Hans C. Fogedby}
\address{
\thanks{Permanent address}
Institute of Physics and Astronomy,
University of Aarhus, DK-8000, Aarhus C, Denmark
\\
and
NORDITA, Blegdamsvej 17, DK-2100, Copenhagen {\O}, Denmark
}
\date{\today}
\maketitle
\begin{abstract}
We discuss the mode spectrum in both the deterministic and
noisy Burgers equations in one dimension. Similar to recent investigations
of vortex depinning in superconductors,
the spectrum is given by a non-Hermitian eigenvalue
problem which is related to a `quantum' problem by a 
complex gauge transformation. The soliton profile in the Burgers equation
serves as a complex gauge field engendering a {\em mode
transmutation} of diffusive modes into
propagating modes and giving rise to a {\em dynamical pinning} 
of localized modes about the solitons.
\end{abstract}
\draft
\begin{multicols}{2}
\narrowtext
The noisy Burgers equation and the related Kardar-Parisi-Zhang (KPZ)
equation provide a continuum description of an intrinsically
nonequilibrium noise-driven system. As such they delimit an interesting
class of systems far from equilibrium. Specifically, the equations
apply to the growth of an interface either due to a random drive
or subject to random environments.

In the case of one spatial dimension, which is our concern here,
the Burgers equation for the local slope $u=\nabla h$ has the form
\cite{Forster}
\begin{eqnarray}
\left(\frac{\partial}{\partial t}-\lambda u\nabla\right)u=
\nu\nabla^2u + \nabla\eta ~,
\label{bur}
\end{eqnarray}
The related KPZ equation for the height $h$ \cite{kpz} is
$\partial h/\partial t=  \nu\nabla^2h + (\lambda/2)(\nabla h)^2 + \eta$.
The damping constant $\nu$ characterizes the linear diffusive
term. The coupling strength $\lambda$ controls the nonlinear growth
or mode coupling term. The noise $\eta$ is spatially 
short-ranged Gaussian white noise correlated according to
$\langle\eta(xt)\eta(00)\rangle = \Delta\delta(x)\delta(t)$
and characterized by the strength $\Delta$.

The stochastic equation (\ref{bur}) and its KPZ version have been studied 
intensively in particular in recent years and much insight 
concerning
the pattern formation and scaling properties
engendered by these equations
has been gained on the
basis of i) field theoretical approaches
\cite{field},
ii) mapping to directed
polymers \cite{Halpin95}, and iii) mapping to the 
asymmetric exclusion model \cite{Derrida98}.

In recent works \cite{Fogedby2} we advanced a Martin-Siggia-Rose
based canonical phase space approach to the noisy Burgers equation
(\ref{bur}). This method applies in the weak noise limit
$\Delta\rightarrow 0$ and replace the stochastic Burgers equation
(\ref{bur}) by two coupled deterministic mean field equations
\begin{eqnarray}
\left(\frac{\partial}{\partial t}-\lambda u\nabla\right)u &&=
\nu\nabla^2 u -\nabla^2p  ~,
\label{mfe1}
\\
\left(\frac{\partial}{\partial t}-\lambda u\nabla\right)p &&=
-\nu\nabla^2 p ~,
\label{mfe2}
\end{eqnarray}
for the slope field $u$ and a canonically conjugate {\em noise field} $p$,
characterizing the noise $\eta$. The field equations derive from a principle
of least action with Hamiltonian density
${\cal H} = p(\nu\nabla^2u + \lambda u\nabla u - (1/2)\nabla^2 p)$ and
determine orbits in a canonical phase space spanned by $u$ and $p$.
Moreover, the action associated with a finite time orbit from $u'$ to
$u$, $S = \int_{0,u'}^{t,u}dtdx(p\partial u/\partial t - {\cal H})$,
provides direct access to the transition probability
$P(u'\rightarrow u,t)\propto\exp[-S/\Delta]$ and associated correlations;
an important aspect which was pursued in 
\cite{Fogedby2}.

On the `zero noise' manifold for $p=0$ the field equation (\ref{mfe1})
reduces to the noiseless Burgers equation \cite{Burgers29}
\begin{eqnarray}
\left(\frac{\partial}{\partial t}-\lambda u\nabla\right)u =
\nu\nabla^2 u  ~,
\label{bur2}
\end{eqnarray}
which as a nonlinear evolution equation exhibiting transient pattern
formation has been used to model `turbulence' and for
example galaxy formation \cite{bur}, see also \cite{Fogedby98a}.

In this Letter we discuss two new features associated with the
pattern formation
in the Burgers equation in both the noiseless case (\ref{bur2}) and
the noisy case in terms of 
(\ref{mfe1}) and (\ref{mfe2}). We focus on the 
interplay between localized nonlinear soliton modes and superposed
linear modes and show the soliton-induced i) {\em mode transmutation}
of diffusive modes into propagating modes and 
ii) {\em dynamical pinning} of linear modes about the solitons.
Details will appear elsewhere.

It is a feature of the nonlinear growth terms that the field equations
(\ref{mfe1}) and (\ref{mfe2}) admit nonlinear localized soliton
solutions, in the static case of the kink-like form
\begin{eqnarray}
u_s^\mu = \mu u\tanh[k_s x] ~, ~~~k_s=\frac{\lambda u}{2\nu}, ~~~~\mu=\pm 1 ~.
\label{sol}
\end{eqnarray}
The index $\mu$ labels the {\em right hand} soliton
for $\mu = 1$ with $p_s=0$, also a solution of the damped noiseless
Burgers equation for $\eta =0$; and the noise-excited {\em left hand}
soliton for $\mu = -1$ with $p_s=2\nu u_s$, a solution of the growing
(unstable) noiseless Burgers equation for $\nu\rightarrow -\nu$.
The amplitude-dependent wavenumber $k_s$
sets the inverse soliton length scale.
Noting that the field equations (\ref{mfe1}) and (\ref{mfe2}) are invariant 
under the slope-dependent Galilean transformation
\begin{eqnarray}
x\rightarrow x - \lambda u_0t ~, ~~~~~u\rightarrow u + u_0 ~,
\label{gal}
\end{eqnarray}
propagating solitons are generated by the Galilean boost (\ref{gal}).
Denoting the right and left boundary values by $u_+$ and $u_-$, respectively,
the propagating velocity is given by the soliton condition
\begin{eqnarray}
u_+ +  u_- = -2v/\lambda ~.
\label{con}
\end{eqnarray}

It follows from the quasi-particle representation advanced in
\cite{Fogedby2}, see also \cite{Fogedby1},
that a general interface slope profile $u = u_s + \delta u$ 
at a particular instant can be
represented by a dilute gas of solitons amplitude-matched according to
(\ref{sol}) with superposed linear modes $\delta u$. 
For a configuration consisting of n solitons we then have 
\begin{eqnarray}
u_s = 
\frac{2\nu}{\lambda}\sum_{p=1}^n k_p\tanh |k_p|(x -v_p t-x_p) ~,
\label{sc}
\end{eqnarray}
where we have introduced the mean amplitude of the p-th soliton
$k_p = (\lambda/4\nu)(u_{p+1} - u_p)$, $u_{p+1}$ and $u_p$ are the
boundary values. The velocity of the p-th soliton is
$v_p = -(\lambda/2)(u_{p+1} + u_p)$, $x_p$ is the center of mass, and we
are assuming vanishing boundary conditions $u_1 = u_{n+1} = 0$. Note
that the configuration (\ref{sc}) is only valid at times in between
soliton collisions; the interface changes dynamically subject to the
conservation of energy, momentum, and total area $\int dx~u$ under
the slope profile. The number of solitons, however, is not conserved,
see  \cite{Fogedby2,Fogedby1}. In Fig 1.
we have depicted an n-soliton configuration
\begin{figure}
\begin{picture}(100,120)
\put(-10.0,10.0)
{
\centerline
{
\epsfxsize=7cm
\epsfbox{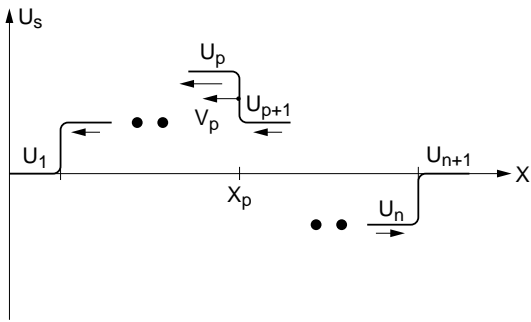}
}
}
\end{picture}
\caption
{
We depict an n-soliton slope configuration of a growing interface.
The p-th soliton moves with velocity 
$v_p = -(\lambda/2)(u_{p+1} + u_p)$, has boundary value $u_+$ and $u_-$,
and is centered at $x_p$. The arrows on the horizontal inter-soliton
segments indicate the propagation of linear modes.
}
\end{figure}

In order to discuss the linear mode spectrum about the soliton
configuration $u_s$ it is convenient to introduce the shifted
noise field $\varphi$
\begin{eqnarray}
p=\nu(u-\varphi) ~,
\label{nf}
\end{eqnarray}
The field equations (\ref{mfe1}) and (\ref{mfe2}) then assume the 
symmetrical form, also discussed in \cite{Fogedby1},
\begin{eqnarray}
&&\left(\frac{\partial}{\partial t}-\lambda u\nabla\right)u=
\nu\nabla^2\varphi ~,
\label{mfea}
\\
&&\left(\frac{\partial}{\partial t}-\lambda u\nabla\right)\varphi=
\nu\nabla^2u ~.
\label{mfeb}
\end{eqnarray}
In the linear Edwards-Wilkinson case \cite{Edwards82} for $\lambda = 0$ 
the field equations readily support
a diffusive mode spectrum of extended growing and decaying modes,
$u\pm\varphi\propto\exp(\mp\nu k^2t)\exp(ikx)$, i.e.,
$u = [A\exp(-\nu k^2t) + B\exp(\nu k^2t)]\exp(ikx)$,  consistent
with the phase space interpretation discussed in 
\cite{Fogedby2}.

Expanding about $u_s$ and the associated noise field $\varphi_s$
($\varphi_s^\mu = \mu u_s^\mu$ for $\mu = \pm 1$),
\begin{eqnarray}
\varphi_s = 
\frac{2\nu}{\lambda}\sum_{p=1}^n|k_p|\tanh|k_p|(x-v_p t-x_p) ~,
\label{sc1}
\end{eqnarray}
$u = u_s + \delta u$ and $\varphi = \varphi_s + \delta\varphi$, the 
superposed linear mode
spectrum is governed  by the coupled non-Hermitian eigenvalue equations
\begin{eqnarray}
&&\left(\frac{\partial}{\partial t}-\lambda u_s\nabla\right)\delta u=
\nu\nabla^2\delta\varphi + \lambda(\nabla u_s)\delta u ~,
\label{eval1}
\\
&&\left(\frac{\partial}{\partial t}-\lambda u_s\nabla\right)\delta\varphi=
\nu\nabla^2\delta u + \lambda(\nabla\varphi_s)\delta u ~.
\label{eval2}
\end{eqnarray}
In the inter-soliton matching regions of constant slope field  
$\nabla u_s = \nabla\varphi_s = 0$ and the equations 
(\ref{eval1}) and (\ref{eval2}) decouple as in the Edwards-Wilkinson
case. Setting $u_s = u$ and 
and searching for solutions of the form
$\delta u, \delta\varphi\propto\exp(-E_kt)\exp(ikx)$ we obtain 
$\delta u \pm \delta\varphi\propto\exp(-E_k^\pm t)\exp(ikx)$, 
i.e.,
$\delta u = [A\exp(-E_k^+t) + B\exp(-E_k^-t)]\exp(ikx)$,
where the complex spectrum characteristic of a non-Hermitian eigenvalue
problem is given by
\begin{eqnarray}
E_k^\pm = \pm\nu k^2 - i\lambda u k ~.
\label{spec}
\end{eqnarray}
Introducing the phase velocity $v  = \lambda u$  the  $\delta u$ mode 
corresponds
to a propagating wave with both a growing and decaying component
\begin{eqnarray}
\delta u\propto(Ae^{-\nu k^2t} + Be^{\nu k^2t})e^{ik(x+vt)} ~.
\label{wave}
\end{eqnarray}

The presence of the nonlinear soliton profile thus gives rise to a
{\em mode transmutation} in the sense
that the diffusive mode in the Edwards-Wilkinson case
is transmuted to a
propagating mode (\ref{wave}) in the Burgers case.
As indicated in Fig. 1 the linear mode propagates to the
left for $u>0$ and to the right for $u<0$. We note in particular that
for a static {\em right hand} soliton $u_\pm =\pm u$ and the mode
propagates towards the soliton center which thus acts like a `sink';
for a static {\em left hand} soliton the situation is reversed,
the mode propagates away from the soliton which in this case plays the
role of a `source'.

In the soliton region the slope field varies over a scale given
by $k_s^{-1}$ and we must address the equations 
(\ref{eval1}) and (\ref{eval2}). 
Introducing the auxiliary variables
\begin{eqnarray}
\delta X^\pm = \delta u \pm \delta\varphi ~,
\label{nv}
\end{eqnarray}
they take the form
\begin{eqnarray}
&&-\frac{\partial\delta X^\pm}{\partial t} = 
\nonumber
\\
&&\pm D\left(\pm\frac{\lambda}{2\nu}u_s\right)\delta X^\pm -
\frac{\lambda}{2}(\nabla u_s \pm \nabla\varphi_s)\delta X^\mp ~,
\label{eval3}
\end{eqnarray}
where $D(\pm\lambda u_s/2\nu)$ is the `gauged' Schr\"{o}dinger operator
\begin{eqnarray}
D\left(\pm\frac{\lambda}{2\nu}u_s\right) = 
-\nu(\nabla\pm\frac{\lambda}{2\nu}u_s)^2
+ \frac{\lambda}{4\nu}u_s^2 - \frac{\lambda}{2}\nabla\varphi_s
\label{ham}
\end{eqnarray}
for the motion of a particle in the potential 
$(\lambda/4\nu)u_s^2 - (\lambda/2)\nabla\varphi_s$ subject to a
gauge field $(\lambda/2\nu)u_s$ given by $u_s$.

In the regions of constant slope field $\nabla u_s = \nabla\varphi_s = 0$,
$u_s = u$, 
$D(\pm\lambda u_s/2\nu)\rightarrow 
-\nu(\nabla\pm\lambda u/2\nu)^2 + (\lambda/4\nu)u^2$ and searching for
solutions of the form $\delta X^\pm\propto\exp(-E_kt)\exp(ikx)$ we
recover the spectrum (\ref{spec}). In the soliton region 
$\nabla\varphi_s^\mu = \mu\nabla u_s^\mu$, 
$\mu = \pm 1$, and one of the
equations (\ref{eval3}) decouples driving the other equation parametrically.

In order to pursue the analysis of (\ref{eval3}) we note that the gauge
field $\lambda u_s/2\nu$ in the Schr\"{o}dinger operator can be absorbed
by means of the gauge or Cole-Hopf \cite{ch} transformation
\begin{eqnarray}
U = \exp\left[-\frac{\lambda}{2\nu}\int dx~ u_s\right] ~.
\label{gauge}
\end{eqnarray}
Using the relation $D(\pm\lambda u_s/2\nu) = U^{\pm 1}D(0)U^{\mp 1}$ we
obtain the Hermitian eigenvalue equations
\begin{eqnarray}
&&-\frac{\partial\delta X^\pm}{\partial t} =
\nonumber
\\
&&\pm U^{\pm 1}D(0)U^{\mp 1}\delta X^\pm -
\frac{\lambda}{2}(\nabla u_s \pm \nabla\varphi_s)\delta X^\mp ~,
\label{eval4}
\end{eqnarray}
which are readily analyzed in terms of the spectrum of $D(0)$ discussed
in \cite{Fogedby98a}.

The exponent or generator in the gauge transformation  (\ref{gauge}) 
samples the area under the slope profile $u_s$ up to the point $x$.
For $x\rightarrow\infty$ $U\rightarrow\exp[-\lambda M/2\nu]$, where
$M = \int dx~ u_s$ is the total area; according to (\ref{bur})
or (\ref{mfe1}) $M$ is a conserved quantity. In terms of the height
field $h$, $u = \nabla h$, $M = h(+L) - h(-L)$ is equal to the height
offset across a system of size $L$, i.e., a conserved quantity under 
growth. Inserting the soliton profile $u_s$ (\ref{sc}) the transformation 
$U$ factorizes in contributions from local solitons, i.e.,
\begin{eqnarray}
U = \prod_{p=1}^n U_p^{\text{sign} k_p}~,~
U_p = \cosh^{-1} k_p(x-v_p t -x_p) ~.
\label{gauge2}
\end{eqnarray}
Focusing on a particular soliton contribution to the interface
with boundary values $u_+$ and $u_-$ and for convenience located
at $x_p = 0$ the analysis is most easily organized by first performing
a Galilean transformation (\ref{gal}) to a local rest frame by shifting
the slope field by $(u_+ + u_-)/2$, corresponding to the velocity
given by (\ref{con}). The static soliton is then given by (\ref{sol}),
$u_\pm = \pm u$, and
\begin{eqnarray}
D(0) = -\nu\nabla^2 +\nu k_s^2\left[1 - 2/\cosh^2k_sx\right]~,
\label{ham2}
\end{eqnarray}
describing the motion of a particle in the attractive 
potential
$-2\nu k_s^2/\cosh^2k_sx$ whose spectrum is known.

Denoting the eigenvalue problem $D(0)\Psi_n = \Omega_n\Psi_n$
the spectrum of $D(0)$ is
composed of
a zero-energy $\Omega_0  = 0$ localized state
$\Psi_0\propto\cosh^{-1} k_sx$, yielding the soliton
translation mode lifting the broken
translational symmetry, and a band $\Psi_k\propto\exp(ik_sx)s_k(x)$
of extended phase-shifted
scattering modes with energy 
\begin{eqnarray}
\Omega_k = \nu(k^2 + k_s^2) ~.
\label{spec3}
\end{eqnarray}
$s_k(x) = (k+ik_s\tanh k_sx)/(k-ik_s)$ is a modulation of the plane
wave state; for $x\rightarrow\infty$
$s_k(x)\rightarrow\exp(i\delta_k)$, where $\delta_k$ is the
phase shift of the wave.

Inserting $\nabla\varphi_s^\mu = \mu\nabla u_s^\mu$ the fluctuations
\begin{eqnarray}
\delta\tilde X^\pm = U_p^{\mp\mu}\delta X^\pm~~, ~~ U_p = \cosh^{-1}k_sx
\label{fluc1}
\end{eqnarray}
then satisfy the Hermitian eigenvalue equations
\begin{eqnarray}
-\frac{\partial\delta\tilde X^\pm}{\partial t} =
\pm D(0)\delta\tilde X^\pm -
\nu(\mu\pm 1)\delta\tilde X^\mp ~,
\label{eval5}
\end{eqnarray}
which decouple and are analyzed by expanding $\delta\tilde X^\pm$ on
the eigenstates $\Psi_n$. For the {\em right hand} soliton for $\mu = +1$
and focusing on the plane wave component, ignoring phase shift effects,
we obtain in particular the fluctuations
\begin{eqnarray}
&&\delta X^+ = (Ae^{-\Omega_kt} +
Be^{\Omega_kt})e^{ikx}\cosh^{-1}k_sx ~,
\label{sol1}
\\
&&\delta X^- = B\frac{\Omega_k}{\nu}e^{\Omega_kt}e^{ikx}\cosh k_sx
~.
\label{sol2}
\end{eqnarray}
For real $k$ the modes $\delta X^\pm$ are diffusive and the spectrum
$\Omega_k = \nu(k^2 + k_s^2)$ exhibits a gap  $\nu k_s^2$ proportional
to the soliton amplitude squared. Moreover, the gauge transformation $U_p$
gives rise to a spatial modulation of the plane wave form which allows us
to extend the spectrum by an analytical continuation in the wavenumber
$k$. 
In particular by setting $k\rightarrow k\mp ik_s$ for $\delta X^\pm$
and noting that $\delta X^\pm$ decouple for $x\gg k_s^{-1}$ we have
$\Omega_k\rightarrow\nu k^2\mp 2i\nu kk_s$ and 
$\exp(ikx)\cosh^{\mp 1}k_s\rightarrow\text{const}$ and we achieve a
matching to the extended propagating modes in the inter-soliton regions.
A similar analysis applies to the {\em left hand} soliton for
$\mu = -1$ and the gauge transformation (\ref{gauge2}) allows for
a complete analysis of the linear fluctuation spectrum about 
the multi-soliton configuration $u_s$.

The last issue we wish to address is the fluctuation spectrum in the
noiseless case, extending the analysis in \cite{Fogedby98a}. In this
case we only have the {\em right hand} soliton for $\mu = +1$ and the
soliton and associated fluctuations lie on the zero-noise manifold
for $p = 0$. There is no coupling to the `noisy' modes 
i.e., $u = \varphi$, $\delta u = \delta\varphi = \delta X^+/2$, and
$\delta X^- = 0$, and the fluctuations are given by (\ref{sol1})
for $B = 0$ (ignoring phase shift effects)
\begin{eqnarray}
\delta u=\delta X^+/2 \propto e^{-\Omega_kt}e^{ikx}\cosh^{-1} k_sx ~.
\label{sol3}
\end{eqnarray}
It is an essential feature of the non-Hermitian eigenvalue
problem (\ref{eval3}) characterizing noisy nonequilibrium growth,
and in the noiseless case of transient growth 
yielding
(\ref{sol3}) that the real spectrum (\ref{spec3}) can be extended 
into the complex eigenvalue
plane. This is due to the envelope $\cosh^{-1}k_s x$ which gives rise
to a spatial fall off. Setting $k\rightarrow k + i\kappa$, 
$\Omega_k\rightarrow E_{k,\kappa}$, where
\begin{eqnarray}
E_{k,\kappa} = \nu (k^2 + k_s^2 - \kappa ^2) + 2i\nu k\kappa ~.
\label{spec2}
\end{eqnarray}
For $|\kappa|<k_s$ we have a band of localized fluctuations 
{\em dynamically pinned} to the soliton. The modes are exponentially 
damped with a damping constant given by the real part of $E_{k,\kappa}$,
$\text{Re} E_{k,\kappa}=\nu(k_s^2-\kappa^2)$. The imaginary part 
of $E_{k,\kappa}$,
$\text{Im} E_{k,\kappa}=2\nu k\kappa$, combined with the phase $ikx$ 
yields a propagating
wave with phase velocity $2\nu\kappa$, finally the spatial range
of the mode is given by $(k_s-\kappa)^{-1}$. For $\kappa = 0$ the
spectrum is real, the phase velocity vanishes, the range is $k_s^{-1}$,
and the localized mode is symmetric and  purely diffusive with  a gap $k_s^2$.
For $\kappa = k_s$, the borderline case, the fluctuations are extended
in space and purely propagating with a gapless spectrum $\nu k^2$. For
intermediate $\kappa$ values the modes are propagating with localized
envelopes. In Fig. 2
we have depicted the complex eigenvalue spectrum.
\begin{figure}
\begin{picture}(100,150)
\put(-10.0,10.0)
{
\centerline
{
\epsfxsize=7cm
\epsfbox{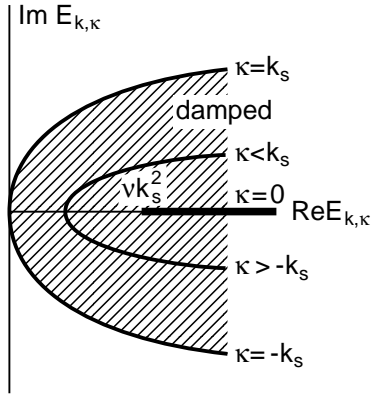}
}
}
\end{picture}
\caption
{
We depict the the complex eigenvalue spectrum for the damped modes
in the noiseless Burgers equation. On the boundaries $\kappa = \pm k_s$
the modes are extended and propagating. The shaded area indicates
propagating modes with localized envelopes. For $\kappa = 0$
the mode is symmetrical and purely diffusive.
}
\end{figure}

In this Letter we have analyzed two aspects of the interplay between
linear superposed modes and nonlinear soliton excitations in the noiseless
and noisy Burgers equations describing transient and stationary 
nonequilibrium growth, respectively. Both aspects are intimately related to the 
non-Hermitian character of the eigenvalue problem. 
The first aspect is a
linear mode transmutation where the diffusive non-propagating modes
in the linear Edwards-Wilkinson case owing to the solitons are 
transmuted to propagating
extended modes in the nonlinear Burgers case. The second  aspect is
a dynamical pinning of a band of localized modes to the solitons.
We finally note that similar aspects
are also encountered in recent work on the transverse Meissner effect and 
flux pinning in superconductors. Here the uniform gauge field is given by
the transverse magnetic field wheras in our case the nonlinear
soliton profile provides the nonuniform gauge field.

Discussions with A. Svane, K. M\o lmer, J. Hertz, B. Derrida, M. L\"{a}ssig,
J. Krug and G. Sch\"{u}tz are gratefully acknowledged.

\end{multicols}

\begin{references}
\bibitem{Forster}
D. Forster, D.~R. Nelson, and M.~J. Stephen, Phys. Rev. Lett. {\bf 36},  867
(1976);
D. Forster, D.~R. Nelson, and M.~J. Stephen, Phys. Rev. A {\bf 16},  732
(1977).

\bibitem{kpz}
M. Kardar, G. Parisi, and Y.~C. Zhang, 
Phys. Rev. Lett. {\bf 56},  889  (1986);
E. Medina, T. Hwa, M. Kardar, and Y.~C. Zhang, Phys. Rev. A {\bf 39},  3053
(1989).

\bibitem{field}
T. Hwa and E. Frey, Phys. Rev. A {\bf 44},  R7873  (1991);
U.~C. T\"{a}uber and E. Frey, Phys. Rev. E {\bf 51},  6319  (1995);
E. Frey, U.~C. T\"{a}uber, and T. Hwa, Phys. Rev. E {\bf 53},  4424  (1996);
E. Frey, U.~C. T\"{a}uber, and H.~K. Janssen, Europhys. Lett {\bf 47},  14
(1999);
M. L\"{a}ssig, Nucl. Phys. B {\bf 448},  559  (1995);
M. L\"{a}ssig, Phys. Rev. Lett. {\bf 80},  2366  (1998).

\bibitem{Halpin95}
T. Halpin-Healy and Y.~C. Zhang, Phys. Rep. {\bf 254},  215  (1995).

\bibitem{Derrida98}
B. Derrida and J.~L. Lebowitz, Phys. Rev. Lett. {\bf 80},  209  (1998).

\bibitem{Fogedby2}
H.~C. Fogedby, Phys. Rev. E {\bf 59},  5065  (1999);
H.~C. Fogedby, Phys. Rev. E {\bf 60},  4950  (1999);
H.~C. Fogedby, cond-mat/0005027.

\bibitem{Burgers29}
J.M.Burgers, Proc. Roy. Neth. Acad. Soc. {\bf 32},  414,643,818  (1929).

\bibitem{bur}
P. Saffman, {\em Topics in Nonlinear Physics, ed. N.J. Zabusky} 
(Springer, New York, 1968);
Y. Zeldovitch, Astron. Astrophys {\bf 8},  84  (1972).

\bibitem{Fogedby98a}
H.~C. Fogedby, Phys. Rev. E {\bf 57},  2331  (1998).

\bibitem{Fogedby1}
H.~C. Fogedby, A.~B. Eriksson, and L.~V. Mikheev, Phys. Rev. Lett. {\bf 75},
1883  (1995);
H.~C. Fogedby, Phys. Rev. E {\bf 57},  4943  (1998);
H.~C. Fogedby, Phys. Rev. Lett. {\bf 80},  1126  (1998).

\bibitem{Edwards82}
S.~F. Edwards and D.~R. Wilkinson, Proc. Roy. Soc. London A {\bf 381},  17
  (1982).

\bibitem{ch}
J.D.Cole, Quart. Appl. Math. {\bf 9},  22  (1951);
E. Hopf, Comm. Pure Appl. Math. {\bf 3},  201  (1950).

\bibitem{hat}
N. Hatano and D.~R. Nelson, Phys. Rev. Lett. {\bf 77},  570  (1996);
N. Hatano and D.~R. Nelson, Phys. Rev. E {\bf 56},  8651  (1997).

\end{references}
\end{document}